\documentclass[aps, prl, twocolumn, superscriptaddress, longbibliography]{revtex4-2}

\usepackage{graphicx}
\usepackage{amsmath,amsthm,amssymb}
\usepackage{dcolumn}
\usepackage{bm}
\usepackage{hyperref}
\usepackage{orcidlink}
\usepackage{tabularx}
\usepackage{array}
\usepackage{bm}
\usepackage{float}

\usepackage{glossaries}

\newacronym{INS}{INS}{Inelastic neutron scattering}

\begin{document}

\preprint{APS/123-QED}

\title{Altermagnetic and dipolar splitting of magnons in FeF$_2$}

\author{J. Sears\orcidlink{0000-0001-6524-8953}}
\email{jsears@bnl.gov}
\affiliation{Condensed Matter Physics and Materials Science Division, Brookhaven National Laboratory, Upton, New York 11973-5000, USA}

\author{V. O. Garlea\orcidlink{0000-0002-5322-7271}}
\affiliation{Neutron Scattering Division, Oak Ridge National Laboratory, Oak Ridge, Tennessee 37831, USA}

\author{D. Lederman\orcidlink{0000-0001-8423-5138}}
\affiliation{Department of Physics, University of California, Santa Cruz, California 95064, USA}

\author{J. M. Tranquada\orcidlink{0000-0003-4984-8857}}
\affiliation{Condensed Matter Physics and Materials Science Division, Brookhaven National Laboratory, Upton, New York 11973-5000, USA}

\author{I. A. Zaliznyak\orcidlink{0000-0002-8548-7924}}
\email{zaliznyak@bnl.gov}
\affiliation{Condensed Matter Physics and Materials Science Division, Brookhaven National Laboratory, Upton, New York 11973-5000, USA}

\date{\today}

\begin{abstract}
FeF$_2$ is a prototypical rutile antiferromagnet recently proposed as an altermagnet, with a magnetic symmetry that permits spin-split electronic bands and chiral magnons. Using very-high-resolution inelastic neutron scattering on a single crystal of FeF$_2$, we show that the dominant source of magnon splitting is in fact the long-range dipolar interaction rather than altermagnetic exchange terms. At momenta where the dipolar splitting vanishes, we observe additional broadening due to altermagnetic chiral splitting and estimate this splitting to be $\sim 35~\mu$eV. Polarized measurements further reveal that, where dipolar splitting is present, the chiral magnon modes become mixed and the resulting modes are predominantly linearly polarized, with at most a small chiral component. These findings highlight the significant effect of dipolar interactions on magnon chirality, particularly when altermagnetic interactions are weak.
\end{abstract}

\maketitle

{\it Introduction}. %
Altermagnets are a recently recognized class of materials in which magnetic symmetry breaks Kramers degeneracy in the absence of a net magnetic moment \cite{smej22a,smej22b}. Altermagnetic spin splitting of electronic bands has been predicted in a wide range of compounds and has now been experimentally detected by angle-resolved photoemission spectroscopy in a small but growing number of materials \cite{krempasky2024_mnte_arpes, lee2024_mnte_arpes, Reimers2024_crsb_arpes, Fedchenko2024_ruo2film_arpes}. The same symmetry condition responsible for electronic band splitting also allows splitting of the magnetic collective excitations into modes of opposite chirality \cite{Andreev_1980, smejkal2023_chiralmag,mcclarty2025_neutron}. Such chiral magnons can, in principle, be detected by inelastic neutron scattering, and experimental searches are currently under way. To date, altermagnetic magnon splitting has been reported in the semiconductors MnTe, Fe$_2$O$_3$ and CrSb \cite{liu2024_mnte_neutron,sun2025_fe2o3_neutron,singh2025_crsb_neutron}. While initial neutron measurements did not detect any splitting in the insulating antiferromagnet MnF$_2$ \cite{morano2025_mnf2}, magnon chirality has now been observed in this material \cite{faure2025_mnf2_neutron}. \\

\begin{figure}
\includegraphics[width=1\columnwidth]{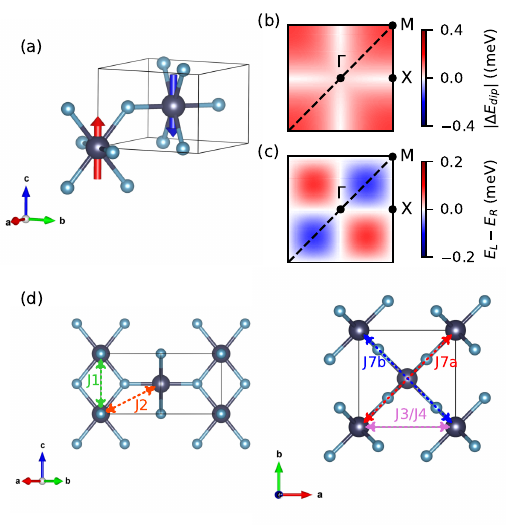}
\caption{Magnetic order and interactions in FeF$_2$. (a) Antiferromagnetic ground state. (b) Modeled momentum-dependence of dipolar splitting in the $(H,K,0)$ plane, showing a maximum at the Brillouin zone boundary. The color scale represents the magnitude of the splitting induced by the dipolar interaction. (c) Modeled momentum-dependence of altermagnetic splitting in the $(H,K,0)$ plane, with a maximum halfway between $\Gamma$ and the Brillouin zone boundary. The color scale represents the difference in energy between the two chiral magnons with altermagnetic anisotropy $J_{7a}-J_{7b}=0.01$~meV. The dashed line shows the scattering plane measured in this experiment. (d) Schematic of the interactions in the model Hamiltonian used to fit the spin wave dispersions. Here $J_3$ couples atoms separated by $(1,0,0)$, while  $J_4$ has a $c$ axis component and connects atoms offset by $(1,0,1)$ in units of the crystallographic unit cell. $J_5$ and $J_6$ were not refined and are omitted from this schematic, but are listed in Table \ref{table:Parameters}. Crystal structure images throughout the manuscript were produces using VESTA \cite{Momma2011}.}
\label{fig1}
\end{figure}

The iron analog, FeF$_2$, possesses the same rutile crystal structure and magnetic ordering as MnF$_2$ [Fig.\ref{fig1}(a)] and so has the symmetry requirements for altermagnetism. Both materials are well known antiferromagnets whose magnetic order and excitations have been studied in the past \cite{erickson1953_mf2_diff,Hutchings1970_fef2_spinwave}. FeF$_2$ is also highly insulating and has an ordering temperature $T_{N} \approx 78 $~K, comparable to that of MnF$_2$. The main difference is the $3d^6$ electronic configuration of Fe, which leads to much stronger single-ion anisotropy compared to the half-filled $d$ shell of Mn. 
Comparing these two materials serves as an interesting test of the effects of the anisotropy on theoretically predicted altermagnetic phenomenology. \\

We have therefore collected a detailed inelastic neutron scattering data set characterizing the magnon dispersion in FeF$_2$. We observe magnon splitting with a size comparable to that previously reported in MnF$_2$. 
We further determine that FeF$_2$ is in a regime where dipolar splitting dominates over altermagnetic effects on the magnon bands, and the same is expected to hold for MnF$_2$. As a result, over most of reciprocal space the two chiral modes are mixed and the magnons appear to be linearly polarized. This is consistent with the results of Ref. \onlinecite{faure2025_mnf2_neutron}, where only a small chiral effect was observed. Our modeling identifies regions of reciprocal space where the dipolar mixing vanishes and magnon chirality is expected to be restored. \\

{\it Methods}. %
\gls*{INS} measurements were performed on the HYSPEC spectrometer \cite{Winn_EJC2015,Zaliznyak_JPhys2017} at the Spallation Neutron Source, Oak Ridge National Laboratory, using a 3.5~g single crystal of FeF$_2$ grown at the University of California, Santa Barbara, of the same origin as those used in previously reported measurements \cite{belanger1987_neutronfef2, BELANGER19831095}. The sample was mounted on an aluminum sample holder with the crystallographic $(H,H,L)$ plane in the horizontal scattering plane. The material has a rutile crystal structure (space group \#136 $P4_2/mnm)$, with lattice constants $a=b=4.69$~\AA, $c=3.31$~\AA; wavevectors are specified in reciprocal lattice units (rlu) of $(2\pi/a,2\pi/b,2\pi/c)$. There are two iron atoms in the unit cell, with moments pointing in opposite directions in the ordered phase [Fig.\ref{fig1}(a)].\\

Unpolarized \gls*{INS} measurements were collected using an incident energy of 13~meV, with flat PG monochromator and chopper frequency of 420 Hz, providing an energy resolution [full width at half maximum (FWHM)] of $\approx 0.35$~meV at the elastic position and $\lesssim 0.15$~meV at energy transfers $E \gtrsim 9$~meV. The data set was collected by rotating the sample through 176$^{\circ}$ in 1$^{\circ}$ steps, with a counting time of $\sim 5$~min per step. Data were collected at two different detector bank positions, covering scattering angles up to $92^{\circ}$, in order to increase
reciprocal-space coverage.\\

The polarized measurements were performed using a vertically focused Heusler crystal monochromator and a supermirror analyzer \cite{Zaliznyak_JPhys2017}, with a Fermi chopper frequency of 300 Hz. The sample was rotated through an angle of 39$^{\circ}$ in 1$^{\circ}$ steps. At each position, the spin-flip and non-spin-flip scattering channels were measured by reversing the polarization of the incident neutron beam using a Mezei spin flipper, with a counting time of $\sim 10$~min per channel. A flipping ratio of 16.5 was measured on the $(110)$ nuclear Bragg peak.\\

\begin{figure}
\includegraphics[width=1\columnwidth]{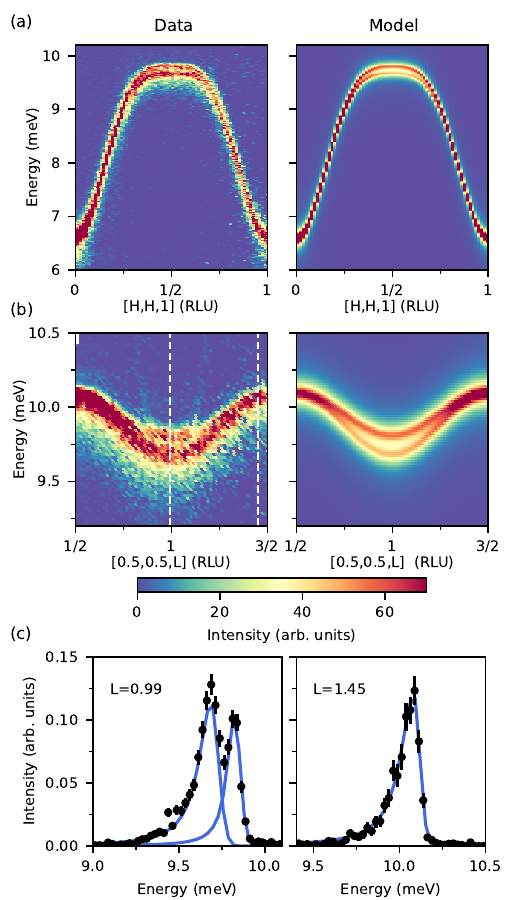}
\caption{Dipolar splitting in FeF$_2$. (a) Dipolar splitting along the $(H,H,1)$ and (b) $(0.5,0.5,L)$ reciprocal space directions. The left panels show experimental data and the right panels show the results of modeling. The vertical white lines indicate the $L$ values for which the line scans are shown in (c). (c) Fits to cuts at $(0.5,0.5,0.99)$ and $(0.5,0.5,1.45)$ with half-Lorentzian, half-Gaussian peak profile.  }
\label{fig2}
\end{figure}

\begin{figure}
\includegraphics[width=1\columnwidth]{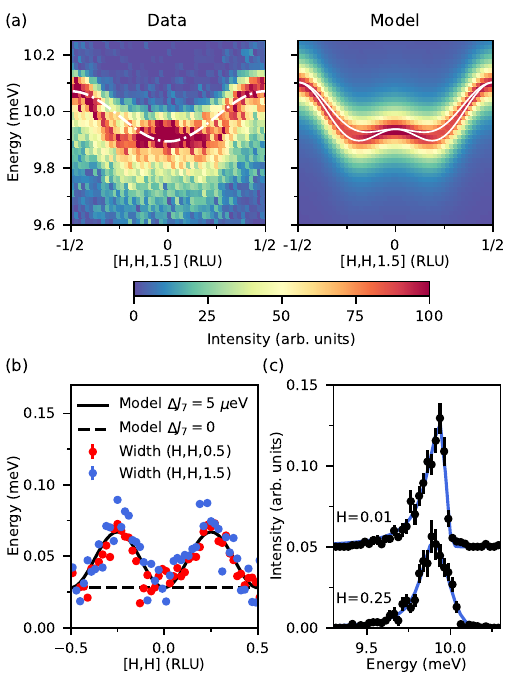}
\caption{Altermagnetic peak broadening at half-integer $L$. (a) Momentum–energy map along the $(H,H,1.5)$ direction. The left panel shows experimental data and the right panel shows the results of modeling. The dash–dot line in the left panel shows the modeled dispersion without the $J_7$ terms, while the solid lines in the right panel show the modeled magnon dispersion energies. (b) Fitted Gaussian peak widths for data at half-integer $L$ values, compared with the calculated splitting between the magnon bands for an altermagnetic anisotropy $\Delta J_7 \equiv \vert J_{7a}-J_{7b}\vert =5~\mu$eV (momentum-dependent splitting) and $\Delta J_7=0~\mu$eV (zero splitting). An offset of 28~$\mu$eV has been added to the calculated splitting to account for the minimum peak width determined by instrumental resolution. (c) Fitted cuts at selected $H$ values with smaller ($H=0.01$) and larger ($H=0.25$) Gaussian peak widths. The $H=0.01$ cut is offset by a constant value of 0.05 for clarity. }
\label{fig3}
\end{figure}

\begin{figure*}
\includegraphics[width=1.9\columnwidth]{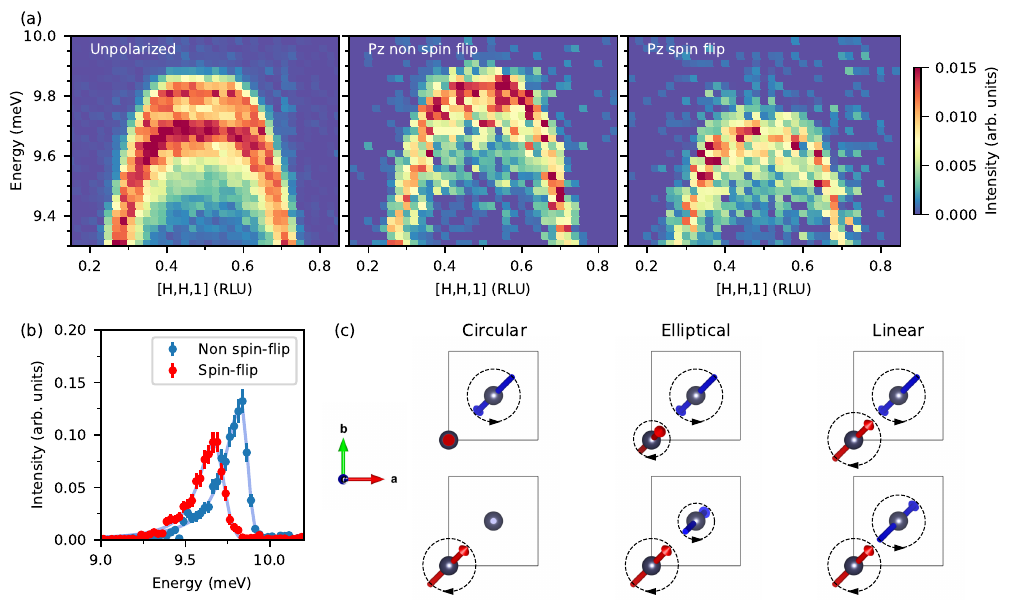}
\caption{Magnon polarization at $L=1$. (a) Experimental momentum-energy maps along $(H,H,1)$ showing predominantly dipolar splitting. The left panel shows data collected with unpolarized neutrons. The center and right panels were collected with vertically polarized neutrons; the center panel shows the non-spin-flip channel, while the right panel shows the spin-flip channel. The separation of the two modes according to polarization channel indicates that these modes are predominantly linearly polarized. (b) Line cuts of the polarized data, integrating over the range $0.4 \le H \le 0.6$. Light blue lines show fits with a mixed Gaussian/Lorentzian peak shape. (c) Schematic illustrating circular chiral magnon modes (left column) and their mixing to form elliptically polarized (center column) and linearly polarized (right column) modes. }
\label{fig4}
\end{figure*}

The data were reduced, normalized, and histogrammed into four-dimensional momentum-energy space using the SHIVER software \cite{saviciShiver2019} with the MDNorm algorithm \cite{MDNorm} implemented in the MANTID package \cite{ARNOLD2014156}. The reciprocal-space coverage in the vertical $(-H,H,0)$ direction was integrated over a range of $\pm 0.05$ reciprocal lattice units to produce a three-dimensional data set limited to the $(H,H,L)$ plane. To extract the magnon energies, constant-wave-vector spectra were fit using a hybrid energy-resolution function consisting of a Lorentzian on the low-energy side of the peak and a Gaussian on the high-energy side, accounting for the asymmetric peak shape. In this way, dispersions along the $(H,H,1)$, $(H,H,0.5)$, $(H,H,1.5)$ and $(0.5,0.5,L)$ directions were obtained and used to refine the spin Hamiltonian. The bin size used for the fitting was 0.02 rlu in both the $[H,H]$ and $L$ directions, and 0.025~meV in energy. The color maps in Figure \ref{fig2} were generated with a bin size of 0.015~rlu in the $[H,H]$ and $L$ directions, and 0.02 meV in energy. The color maps in Figures \ref{fig3} and \ref{fig4} were generated with a bin sizes of 0.02 and 0.04~rlu in the $[H,H]$ and $L$ directions respectively, and 0.025 meV in energy.\\

All modeling was performed using the SUNNY package \cite{dahlbom2025sunnyjljuliapackagespin} which uses linear spin-wave theory to simulate the spin-wave energies and the corresponding neutron scattering intensity. The colormaps of the modeled magnon dispersions were produced using a 0.1 meV Gaussian resolution, and the permittivity as defined within SUNNY implementation of the magnetic dipole interaction was set to 85\% of the free space value to best match the observed size of the dipolar splitting.\\

{\it Results}. %
The inelastic neutron scattering measurements collected at 4~K in the magnetically ordered phase show sharp, dispersive spin-wave excitations between 6 and 10.5 meV over the full range of reciprocal space measured. At most wave vectors only a single mode is observed; however, in some regions a small splitting on the order of 0.1 meV is clearly visible as shown in the left-hand panels of Fig.~\ref{fig2} (a) and (b). This splitting is largest at the Brillouin zone boundary, at positions of the type $(\frac{2n+1}{2}, \frac{2n+1}{2}, m)$ where $n$ and $m$ are integers. This momentum dependence is not consistent with altermagnetic chiral splitting, which arises from anisotropy between the magnetic interaction terms 
\cite{smejkal2023_chiralmag, faure2025_mnf2_neutron} (to be discussed). Chiral splitting of this type is expected to vanish at the Brillouin zone boundary, as shown in Fig.~\ref{fig1}(c). By contrast, splitting due to long-range magnetic dipole interactions [Fig.~\ref{fig1}(b)] is expected to be largest at the Brillouin zone boundary, in agreement with our observations.\\

To clarify the role of altermagnetic interactions, we have modeled the spin-waves using a Hamiltonian including several isotropic Heisenberg interactions, easy-axis single ion anisotropy $D$, and long-range dipole interactions:

\begin{equation}
{\cal H} = \sum_{\langle i, j \rangle_n} J_n \bm{S}_i \cdot \bm{S}_j - \sum_i D (S_i^z)^2 + {\cal H}_{dip} .
\end{equation}
Here $n$ indexes pairs of atoms, as shown in Fig.~\ref{fig1}(d). We consider $n$ from 1 to 7, as done for MnF$_2$ \cite{faure2025_mnf2_neutron}; the associated vectors $\bm{x}_{\langle i, j \rangle_n}$ connecting atomic pairs are listed in Table \ref{table:Parameters}. We included long-range dipolar interactions as implemented in SUNNY \cite{dahlbom2025sunnyjljuliapackagespin}, consistent with previous work on FeF$_2$ \cite{Hutchings1970_fef2_spinwave}. The interactions in the Hamiltonian were optimized by fitting the measured spin wave dispersion in the $(H,H,L)$ plane. \\

We find that including the interactions $D$, $J_1$, $J_2$, $J_4$, and $J_{7a/b}$ already provides a very good fit of the dispersion with a minimal number of parameters. The fitted values are listed in Table~\ref{table:Parameters}, and the corresponding spin-wave simulations shown in the right-hand panels of Fig.~\ref{fig2}(a) and (b) are in good agreement with the data. The parameters $D$ and $J_2$ are the largest and are primarily responsible for the large dispersion at integer $L$ values; we estimate their uncertainties to be approximately 5\% of their refined values. The next-largest term, $J_1$, with an estimated uncertainty of 10\%, accounts for the $\sim 0.5$~meV dispersion along $L$. The $J_4$ and $J_{7a/b}$ terms give rise to the more subtle dispersion at half-integer $L$ values, and we estimate uncertainties of 20\% and 30\%, respectively. \\

\begin{table}
\caption{{
Refined values of the Heisenberg exchange parameters $J_n$ and the vectors $\bm{x}_{\langle i, j \rangle_n}$ connecting the corresponding atomic pairs; $d_n$ is the signed distance between the corresponding magnetic ions, where a negative sign indicates antiparallel moments. }}
\centering
\begin{ruledtabular}
\begin{tabular}{ c c c c  }
 Interaction & Fit value  & $d_n$ (\AA) & $\bm{x}_{\langle i, j \rangle_n}$ \\
\hline
D & 0.82(4) &  - & -  \\
 $J_2$ & 0.46(2) &(-)3.71 & $(\frac{1}{2},\frac{1}{2},\frac{1}{2})$ \\
 $J_1$ & -0.030(3) & 3.31 & $(0,0,1)$ \\
 $J_3$ & - & 4.69 & $(1,0,0)$ \\
 $J_4$ & 0.005(1) & 5.74 & $(1,0,1)$\\
 $J_5$ & - & (-)5.97 & $(\frac{1}{2},\frac{1}{2},\frac{3}{2})$  \\
 $J_6$ & - & 6.62 & $(0,0,2)$  \\
 $J_{7a} + J_{7b}$ & 0.022(7) & - & - \\
 $J_{7a}$ & - & 6.63 & $(1,1,0)$  \\
 $J_{7b}$ & - & 6.63 & $(-1,1,0)$  \\

\end{tabular}
\end{ruledtabular}
\label{table:Parameters}
\end{table}

Introducing additional interactions up to $J_6$ (following the convention of Ref.~\cite{faure2025_mnf2_neutron}) did not improve the fit, and these variables exhibited strong covariance with the refined parameters. In particular, $J_3$ and $J_4$ have very similar effects on the dispersion in this scattering plane, with $J_4$ yielding a slightly better fit. The refined $J_4$ term likely subsumes contributions from $J_3$ and cannot be independently refined. Likewise, the further-neighbor interactions $J_5$ and $J_6$ cannot be disentangled from the refined parameters on the basis of the present data set. We note that the integrated inelastic intensity is well fit using a minimal model including only $D$ and $J_2$ as shown in the End Matter, consistent with our refinement results which indicate that these are the dominant interactions.\\

The first interactions that reflect the reduced symmetry associated with altermagnetism are $J_{7a}$ and $J_{7b}$. Along the $[1,1,0]$ direction, $J_{7a}$ corresponds to the path Fe$_{\uparrow}$–F–Fe$_{\downarrow}$–F–Fe$_{\uparrow}$, whereas $J_{7b}$ follows Fe$_{\downarrow}$–F–Fe$_{\uparrow}$–F–Fe$_{\downarrow}$; these paths involve inequivalent bond angles, resulting in unequal exchange interactions and hence chiral splitting of the magnons \cite{faure2025_mnf2_neutron}. To isolate this effect, we have to examine the role of the dipolar interactions. \\

The $(\frac{1}{2},\frac{1}{2},L)$ cut in Fig.~\ref{fig2}(b) shows that the dipolar splitting varies in magnitude as a function of $L$. In contrast, altermagnetic splitting is expected to be $L$ independent. At integer $L$ values the dipolar splitting dominates, while at half-integer $L$ it is expected to vanish, leaving altermagnetic splitting as the only source of spectral broadening. We therefore examine cuts at half-integer $L$ [Fig.~\ref{fig3}(a)] to search for signatures of altermagnetic splitting. The dispersion is highly sensitive to the inclusion of the $J_7$ terms: when $J_{7a}+J_{7b}\!\approx\!0$, the model yields a sinusoidal dispersion qualitatively different from the square shaped dispersion observed [left-hand panel of Fig.~\ref{fig3}(a)]. Reproducing the measured dispersion requires refining $J_{7a}+J_{7b}$ to $22 \pm 7~\mu$eV. \\

While the magnon dispersion constrains the sum $J_{7a}+J_{7b}$, chiral splitting requires an anisotropy $J_{7a}-J_{7b}\neq 0$ and would manifest as a splitting between magnons of opposite chirality. While no splitting can be definitively resolved, spectra along $(H,H,1.5)$ exhibit a measurable $q$-dependent broadening [Fig.~\ref{fig3}(b)]. Introducing an anisotropy $J_{7a}-J_{7b}=5~\mu$eV into our model yields a calculated chiral splitting that provides a good match to the observed broadening, as shown by the solid line in Fig.~\ref{fig3}(b). These results suggest that the altermagnetic coupling is on the order of that recently found in MnF$_2$ \cite{faure2025_mnf2_neutron}. Although the altermagnetic splitting of $\sim 35~\mu$eV is approximately three times smaller than the dipolar splitting, magnon chirality should be observable in FeF$_2$ in the presence of a magnetic-domain imbalance. \\

The effect of the dipolar interaction is to mix the chiral magnon modes \cite{faure2025_mnf2_neutron,Jin2025_mixing}, such that in regions of reciprocal space where dipolar splitting is present the chiral modes are nearly fully mixed and predominantly linearly polarized. This is illustrated in Fig.~\ref{fig4}, which compares measurements with unpolarized and vertically polarized neutrons (Pz), where the polarization direction is perpendicular to the scattering plane.. The unpolarized data show both modes, whereas the polarized measurements reveal the upper mode in the non-spin-flip channel and the lower mode in the spin-flip channel. Since only fluctuations perpendicular to the neutron polarization axis contribute to spin-flip scattering, this indicates that the upper mode is polarized along the vertical $(-H,H,0)$ direction, while the lower mode is polarized in the horizontal $(H,H,0)$ direction. The ordered magnetic moment lies  along the crystallographic $c$ axis, i.e., $(0,0,L)$ with spin-wave modes corresponding to transverse fluctuations. \\

These observations can be understood by examining the calculated eigenmodes of the spin Hamiltonian. Chiral magnon modes, illustrated in the left column of Fig.\ref{fig4}(c), involve circular precession of only one of the two magnetic moments in the unit cell. The involvement of a single sublattice gives rise to chirality, as the mode carries a net moment. However, such modes are not favored in the presence of dipolar interactions, which couple the two sublattices. The dipolar interaction therefore mixes the chiral modes to form elliptically or linearly polarized modes, constructed from combinations of the chiral modes as shown in Fig.\ref{fig4}(c). 
As shown in the upper row of Fig.\ref{fig4}(c), for spin waves near the Brillouin zone boundary the $[1,1,0]$ components of the two moments in the unit cell tend to cancel, yielding an overall polarization primarily along $[-1,1,0]$. In the lower row, the $[-1,1,0]$ components cancel, resulting in polarization along $[1,1,0]$. The observation of linear polarization at integer $L$ values indicates a roughly equal mixture of the two chiral modes within the experimental sensitivity. We note that a residual degree of chirality is still expected, and indeed has been detected in MnF$_2$ \cite{faure2025_mnf2_neutron}.\\

{\it Conclusions}. %
Our inelastic neutron scattering measurements have enabled a characterization of magnetic excitations in FeF$_2$ with a much higher resolution than previously achieved. This has allowed us to observe the effects of the altermagnetic anisotropy, which we estimate to be comparable in magnitude to that recently reported in MnF$_2$ \cite{faure2025_mnf2_neutron}. Our calculations indicate that the enhanced single-ion anisotropy in FeF$_2$ does not significantly influence magnon chirality. \\

In the present sample, we did not assess the relative populations of the two magnetic domains, which are related by time reversal symmetry; however, a significant domain imbalance has previously been observed or induced in both Mn- and Fe-based materials. It is therefore likely that magnon chirality will be detectable via polarized neutron measurements of the type employed in Ref.~\cite{faure2025_mnf2_neutron}. Our calculations further show that chirality is suppressed at integer $L$ values and restored at half-integer $L$ values, where the dipolar splitting vanishes. These results highlight the important effect of dipolar interactions on magnon chirality and polarization, particularly in insulating materials where longer-range chiral magnetic interactions are weak.

{\it Acknowledgments}. %
We are grateful to Melissa Graves-Brook for assistance with the measurements and to Andrei Savici for help with data reduction. 
Work at Brookhaven is supported by the Office of Basic Energy Sciences, Materials Sciences and Engineering Division, U.S.\ Department of Energy (DOE) under Contract No.\ DE-SC0012704. This research used resources at the Spallation Neutron Source, a DOE Office of Science User Facility operated by Oak Ridge National Laboratory. The beamtime was allocated to HYSPEC on proposal number IPTS-34335. Work at UC Santa Cruz was supported in part by a University of California National Laboratory Fees Research Program, Grant No. L25CR8980.

\section*{Data Availability}
The data that support the findings of this article are openly available on the Zenodo database under the access code 20029957 \cite{repo}.

\clearpage
\appendix
\section{End Matter: Fit to integrated intensity}
\label{Appendix:Fit_intensity}
For a single-mode spin-excitation spectrum, such as the magnon excitations in FeF$_2$ measured here, the first-moment sum rule yields the following expression, which relates the static structure factor $S^{\alpha\alpha}(\bm{Q})$, obtained from the energy-integrated spin-wave intensity with polarization $\alpha$, to the interaction parameters of the spin Hamiltonian \cite{ZaliznyakLee_MNSChapter}
\begin{equation}
\begin{split}
\omega^{\alpha}(\bm{Q}) S^{\alpha\alpha}(\bm{Q}) \propto & - \sum_{\langle i, j, \rangle_n} J_{ij} \langle S^c_i S^c_j \rangle (1-\cos(\bm{Q} \cdot \bm{x}_{i,j})) \\
-& \sum_i D \langle (S_i^c)^2\rangle .
\end{split}
\label{eq2:1st_moment}
\end{equation}
Here $\omega^{\alpha}(\bm{Q})$ is the magnon energy and $S^{\alpha\alpha}(\bm{Q})$ is the momentum-dependent scattered intensity. The first summation runs over spins $i$ within the unit cell and their $n^{\rm th}$-neighbor partners $j$, with $\bm{x}_{ij}$ denoting the real-space vector connecting sites $i$ and $j$; $J_{ij} = J_n$ is the corresponding Heisenberg exchange interaction. The second term represents the contribution from the single-ion anisotropy $D$. For the classical N\'eel-ordered state within the linear spin-wave approximation, transverse spin correlations vanish and are neglected, $\langle (S_i^{a,b})^2 \rangle = \langle S_i^{a,b} S_j^{a,b} \rangle = 0$, while the longitudinal correlations satisfy $\langle (S_i^c)^2 \rangle = S^2$ and $\langle S_i^{c} S_j^{c} \rangle = \pm S^2$, with the sign determined by whether $\bm{x}_{ij}$ connects spins on the same or opposite magnetic sublattices.

\begin{figure}[!b]
\includegraphics[width=\columnwidth]{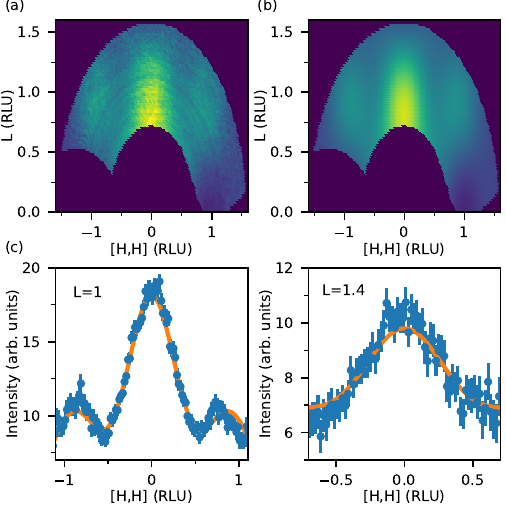}
\caption{ (a) Scattered intensity, integrated over energy and multiplied by magnon energy extracted from the data. (b) Modeled intensity using equation \ref{eq2:1st_moment} including only $D$ and $J_2$. (c) Cuts comparing modeled and measured intensities at two different $L$ values. }
\label{fig5}
\end{figure}

The total measured intensity $S(\bm{Q})$ includes the polarization prefactor
\begin{equation}
1 + \cos^2 \phi = 1 + \frac{(c^* L)^2}{Q^2} ,
\end{equation}
where $\phi$ is the angle between the momentum transfer $\bm{Q}$ and the crystallographic $c$ axis, and $c^*$ is the reciprocal lattice parameter. This prefactor was applied to the right-hand side of Eq.~\eqref{eq2:1st_moment} when fitting $S(\bm{Q})$, accounting for the presence of two spin-wave modes with polarizations along $(-H,H,0)$ (perpendicular to the scattering plane) and along $(H,H,0)$ (within the scattering plane). The first mode carries a polarization factor of $1$, while the second carries a factor of $\cos^2 \phi$ and is therefore suppressed when $\bm{Q}$ lies along $(H,H,0)$.

This analytic form can be compared with the momentum-dependent intensity, integrated over energy and multiplied by the fitted energy of the magnon. Since the magnon dispersion has a large energy gap of $6$~meV this intensity is not strongly affected by elastic contributions. The experimental data processed in this way is shown in Fig.\ref{fig5}(a), and can be reasonably well reproduced using a model including only the largest interactions $D$ and $J_2$ in addition to a phonon term proportional to $\bm{Q}^2$. Including further-range interactions did not improve the fit to the data, and so these interactions could not be refined. This modeled intensity is shown in Fig.\ref{fig5}(b) as well as the reciprocal space cuts shown in Fig.\ref{fig5}(c). \\

This fit to the intensity data allows us to determine the ratio of the $D$ and $J_2$ terms, which we find to be $\sim 2.1$, somewhat larger than the value of $\sim 1.8$ obtained from fitting the dispersion. This discrepancy likely indicates the presence of additional background contributions in the intensity data. Since including further $J$ interactions does not improve the fit, we conclude that these terms are small compared to the dominant terms $D$ and $J_2$, lying within the $\approx 10\%$ systematic uncertainty of the spin wave intensity data, as indicated by the refined $D/J_2$ mismatch noted above. This conclusion is consistent with the dispersion fitting described in the main text. \\

\end{document}